\documentclass[a4paper,conference,twocolumn,10pt]{IEEEtran}

\usepackage[T1]{fontenc}
\usepackage[utf8]{inputenc}
\usepackage{stmaryrd}
\usepackage{amssymb}
\usepackage{amsmath}
\usepackage{amsfonts}
\usepackage{mathrsfs}
\usepackage{bbm}
\usepackage{todonotes}

\usepackage{color}
\definecolor{darkred}{RGB}{174,0,0}        
\definecolor{darkgreen}{RGB}{0,166,0}      
\definecolor{darkblue}{rgb}{0.00,0.00,0.65}       

\usepackage{graphicx}
\usepackage{epstopdf}
\usepackage{caption}
\usepackage{subcaption}
\usepackage{float}

\usepackage{fullpage}
\usepackage{multicol}
\usepackage{enumitem}
\usepackage{url}

\title{\Large \bf
        GNU Radio Implementation of MALIN:\\
        ``Multi-Armed bandits Learning for Internet-of-things Networks''
}

\author{\IEEEauthorblockN{Lilian Besson\IEEEauthorrefmark{1}, R\'emi Bonnefoi\IEEEauthorrefmark{1}, and Christophe Moy\IEEEauthorrefmark{2}}

\IEEEauthorblockA{\IEEEauthorrefmark{1}CentraleSup\'elec/IETR, CentraleSup\'elec Campus de Rennes, $35510$ Cesson-S\'evign\'e, France\\
Email: \texttt{\{lilian.besson, remi.bonnefoi\}{@}centralesupelec.fr}}

\IEEEauthorblockA{\IEEEauthorrefmark{2}Univ Rennes, CNRS, IETR -- UMR $6164$, $35000$ Rennes, France\\
Email: \texttt{christophe.moy{@}univ-rennes1.fr}}
}

\graphicspath{{pictures/}}
\DeclareGraphicsExtensions{.jpg,.pdf,.mps,.eps,.png}

\newcommand{\UCB}{\ensuremath{\mathrm{UCB}_1}}

\begin{document}

\date{}

\maketitle

\pagenumbering{arabic}
\setcounter{page}{1}

\begin{abstract}

We implement an IoT network the following way: one gateway, one or several intelligent (\emph{i.e.}, learning) objects, embedding the proposed solution,
and a traffic generator that emulates radio interferences from many other objects.
Intelligent objects communicate with the gateway with a wireless ALOHA-based protocol, which does not require any specific overhead for the learning.
We model the network access as a discrete sequential decision making problem, and using the framework and algorithms from Multi-Armed Bandit (MAB) learning, we show that intelligent objects can improve their access to the network by using low complexity and decentralized algorithms, such as \UCB{} and Thompson Sampling.
This solution could be added in a straightforward and costless manner in LoRaWAN networks, just by adding this feature in some or all the devices, without any modification on the network side.

\end{abstract}


\section{Introduction}

The monitoring of large scale systems, such as smart grids and smart cities, requires the development of networks dedicated to Internet-of-Things (IoT) applications.
For instance, Low Power Wide Area Networks (LPWAN) \cite{Raza17}, as LoRaWAN or SigFox, are nowadays deployed in unlicensed bands to handle a large number of objects transmitting a few packets per day or week.
In order to reduce the energy consumption of end-devices, these networks rely on pure ALOHA-based Medium Access (MAC) protocols.

One of the challenges in the design of MAC solutions for the IoT is to design solutions which improve the performance of the network and reduce the Packet Loss Ratio (PLR), without reducing the end-devices battery life.
In particular, many IoT standards operate in unlicensed bands, that is why we have to find solutions that do not increase the PLR due to the interference caused by other standards and networks which share the same band, without coordination.
As this interfering traffic is generated by different standards and networks, it cannot be controlled, and it is not evenly distributed in the different channels.

Multi-Armed Bandit (MAB) algorithms \cite{bubeck2012regret} have been recently proposed as a solution to improve the performance of IoT networks and in particular in LPWAN \cite{Bonnefoi18,Azari18}.
In this paper, we describe the way we implemented a demo where we evaluate MAB algorithms \cite{bubeck2012regret}, used in combination with a pure ALOHA-based protocol (such as the ones employed in LPWAN).
%
%
This demonstration is the first implementation which aims at assessing the potential gain of MAB learning algorithms in IoT scenarios.
Following our recent work \cite{Bonnefoi17}, we propose to model this problem as Non-Stationary\footnote{~Note that non-stationarity only comes from the presence of more than one dynamic object, as the background traffic is assumed independent and identically distributed \emph{i.i.d.}.} MAB.
We suggest to use low-cost algorithms, focusing on two well-known algorithms: a frequentist one (\UCB{}) and a Bayesian one (TS).
We consider the Upper-Confidence Bound (\UCB{}) \cite{Auer}, and the Thompson Sampling (TS) algorithms \cite{Thompson33}. Both algorithms have already been applied with success in the context of wireless decision making, both empirically for Opportunistic Spectrum Access \cite{Jouini},
and more recently for multi-users Cognitive Radio problems \cite{BessonALT18} with a more theoretical approach.

We use a TestBed designed in 2017 by our team SCEE \cite{Bodinier17}, containing different USRP boards \cite{USRPDocumentation}, controlled by a single laptop using GNU Radio \cite{GNURadioDocumentation},
and where the intelligence of each object corresponds to a learning algorithm, implemented as a GNU Radio block \cite{GNURadioCompanionDocumentation} and written in Python or \texttt{C++}.

In our demo, we consider a simple wireless network, consisting of one gateway (radio access point), and a certain interfering background traffic, assumed to be stationary (\emph{i.i.d.}), which is generated by end-devices communicating in other networks.
Some dynamic intelligent objects (end-user or autonomous objects) try to communicate with the gateway, with a low-overhead protocol. This communication can be done in different channels which are also shared by devices using other networks.
Once the gateway receives a packet transmitted by a dynamic device in one channel, it transmits back to it an acknowledgement in the same channel, after a fixed-time delay, as it is done in the LoRaWAN standard.
This \emph{ACK} allows the device to learn about the channel quality and thus, to use learning algorithms for the purpose of best channel selection.

In this demo, we can generate scenarios with different parameters (number of channels, interfering traffic, etc) in order to evaluate the performance of learning in various settings.
Moreover, we compare the performance of learning with that of the random uniform access to channels, which is the current state-of-the-art of commercial LPWAN solutions.
This allows to check that in case of uniform traffic, when there is nothing to learn, the intelligent objects at least do not reduce their successful communication rate in comparison to the naive objects.
This also shows that in case of non-uniform stationary traffic, the MAB learning algorithms indeed help to increase the global efficiency of the network by improving the success rate of the intelligent objects.

The rest of this paper is organized as follows. The system model is introduced in Section 2. In Section 3, we describe more formally both the \UCB{} and the TS algorithms. Our implementation is presented in Section 4, and results are given in Section 5.

\section{System Model}

\begin{figure}[!t]
    \centering
    \includegraphics[width=0.90\columnwidth]{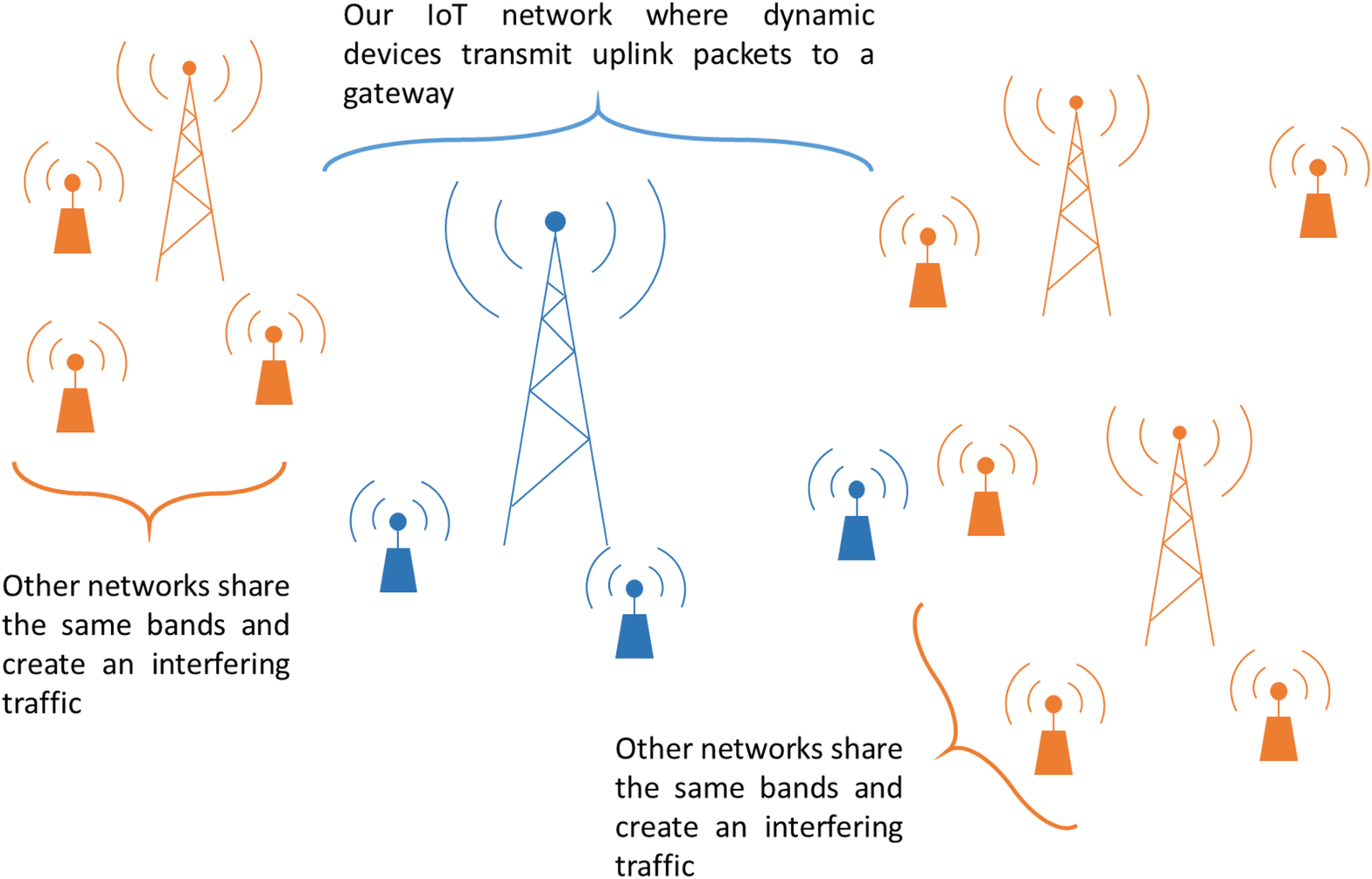}
    \caption{In our system model, some dynamic devices transmit packets to a gateway and suffer from the interference generated by neighboring networks.}
    \label{fig:system_model1}
\end{figure}

We consider the system model presented in Figure~\ref{fig:system_model1}, where a set of object sends uplink packets to the network gateway, in the $433.5\;\mathrm{MHz}$ ISM band.
The communication between IoT devices and this gateway is done through a simple pure ALOHA-based protocol where devices transmit uplink packets of fixed duration whenever they want.
The devices can transmit their packets in $K\geq 1$ channels (\emph{e.g.}, $K=4$). In the case where the gateway receives an uplink in one channel, it transmits an acknowledgement to the end-device in the same channel, after a fixed delay (of $1$ s).

These communications operate in unlicensed ISM bands and, consequently, suffer from interference generated by uncoordinated neighboring networks. This interfering traffic is uncontrolled, and can be unevenly distributed over the $K$ different channels.

%

We consider the network from the point of view of one end-user. Every times the end-user has to communicate with the gateway,
it has to choose one channel (at each transmission $t \geq 1, t \in \mathbb{N}$), denoted as $C(t) = k \in\{1,\dots,K\}$.
Then, the end-users starts waiting in this channel $C(t)$ for an acknowledgement sent by the gateway.
Before sending another message (\emph{i.e.}, at time $t+1$), the end-user knows if it received or not this \emph{ACK} message.
For this reason, selecting channel (or arm) $k$ at time $t$ yields a (random) feedback, called a \emph{reward}, $r_k(t) \in \{0,1\}$, being $0$ if no \emph{ACK} was received before the next message, or $1$ if \emph{ACK} was successfully received.
The goal of the end-user is to minimize its packet loss ratio, or equivalently, it is to maximize its cumulative reward,
$r_{1 \dots T} := \sum_{t = 1}^T r_{C(t)}(t),$
as it is usually done in MAB problems \cite{Thompson33,Robbins52,LaiRobbins85}.

This problem is a special case of the so-called ``stochastic'' MAB, where the sequence of rewards drawn from a given arm $k$ is assumed to be  \emph{i.i.d.}, under some distribution $\nu_k$, that has a mean $\mu_k$. Several types of reward distributions have been considered in the literature, for example distributions that belong to a one-dimensional exponential family (\emph{e.g.}, Gaussian, Exponential, Poisson or Bernoulli distributions).

Rewards are binary in our model, and so we consider only Bernoulli distributions, in which $r_k(t) \sim \mathrm{Bern}(\mu_k)$, that is, $r_k(t) \in \{0,1\}$ and $\mathbb{P}(r_k(t) = 1) = \mu_k \in [0,1]$.
Contrary to many previous work done in the CR field (\emph{e.g.}, Opportunistic Spectrum Access),
the reward $r_k(t)$ does \emph{not} come from a sensing phase before sending the $t$-th message, as it would do for any ``listen-before-talk'' model. Rewards come from receiving an acknowledgement from the gateway, between the $t$-th and $t+1$-th messages.

The problem parameters $\mu_1,\dots,\mu_K$ are of course unknown to the end-users, so to maximize its cumulated reward, it must learn the distributions of the channels, in order to be able to progressively focus on the best arm (\emph{i.e.}, the arm with largest mean).
This requires to tackle the so-called \emph{exploration-exploitation dilemma}: a player has to try all arms a sufficient number of times to get a robust estimate of their qualities, while not selecting the worst arms too many times.


\section{MAB Algorithms}

Before discussing the relevance of a MAB model for our IoT application, we present two bandit algorithms, \UCB{} and Thompson Sampling,
which are both known to be efficient for stationary \emph{i.i.d.} rewards and are shown to be useful in our setting (in Sec.~\ref{sec:results}).

\subsection{The \UCB{} Algorithm}\label{sub:UCB}

A naive approach could be to use an empirical mean estimator of the rewards for each channel, and select the channel with highest estimated mean at each time;
but this greedy approach is known to fail dramatically \cite{LaiRobbins85}. Indeed, with this policy, the selection of arms is highly dependent on the first draws:
if the first transmission in one channel fails and the first one on other channels succeed, the end-user will \emph{never} use the first channel again, even it is the best one (\emph{i.e.}, the most available, in average).

Rather than relying on the empirical mean reward, Upper Confidence Bounds algorithms
instead use a \emph{confidence interval} on the unknown mean $\mu_k$ of each arm,
which can be viewed as adding a ``bonus'' exploration to the empirical mean.
They follow the ``\emph{optimism-in-face-of-uncertainty}'' principle: at each step, they play according to the best model,
as the statistically best possible arm (\emph{i.e.}, the highest upper confidence bound) is selected.

More formally, for one end-user, let $N_k(t) = \sum_{\tau=1}^t \mathbbm{1}(C(\tau) = k)$ be the number of times channel $k$ was selected up-to time $t \geq 1$.
The empirical mean estimator of channel $k$ is defined as the mean reward obtained by selecting it up to time $t$, $\widehat{\mu_k}(t) = 1 / N_k(t) \sum_{\tau=1}^t r_k(\tau) \mathbbm{1}(C(\tau) = k) $.
For \UCB, the \emph{confidence} term is $B_k(t) = \sqrt{\alpha \log(t) / N_k(t)}$,
giving the upper confidence bound $U_k(t) = \widehat{\mu_k}(t) + B_k(t)$, which is used by the end-user to decide the channel for communicating at time step $t+1$: $C(t+1) = \arg\max_{1\leq k \leq K} U_k(t)$.
\UCB{} is called an \emph{index policy}.

The \UCB{} algorithm uses a parameter $\alpha > 0$, originally $\alpha$ was set to $2$ \cite{Auer}, but empirically $\alpha = 1/2$ is known to work better (uniformly across problems), and $\alpha > 1/2$ is advised by the theory \cite{bubeck2012regret}.
%
In our model, every dynamic end-user implements its own \UCB{} algorithm, \emph{independently}.
For one end-user, the time $t$ is the total number of sent messages from the beginning, as rewards are only obtained after a transmission.

\subsection{Thompson Sampling}

Thompson Sampling \cite{Thompson33} was introduced early on, in $1933$ as the very first bandit algorithm, in the context of clinical trials (in which each arm models the efficacy of one treatment across patients). Given a prior distribution on the mean of each arm, the algorithm selects the next arm to draw based on samples from the \emph{conjugated} posterior distribution, which for Bernoulli rewards is a Beta distribution.

A Beta prior $\mathrm{Beta}(a_k(0)=1,b_k(0)=1)$ (initially uniform) is assumed on $\mu_k \in [0, 1]$, and at time $t$ the posterior is $\mathrm{Beta}(a_k(t),b_k(t))$.
After every channel selection, the posterior is updated to have $a_k(t)$ and $b_k(t)$ counting the number of successful and failed transmissions made on channel $k$.
So if the \emph{ACK} message is received, $a_k(t+1) = a_k(t) + 1$, and $b_k(t+1) = b_k(t)$, otherwise $a_k(t+1) = a_k(t)$, and $b_k(t+1) = b_k(t) + 1$.
Then, the decision is done by \emph{sampling} an \emph{index} for each arm, at each time step $t$, from the arm posteriors: $X_k(t) \sim \mathrm{Beta}(a_k(t), b_k(t))$, and the chosen channel is simply the channel $C(t+1)$ with highest index $X_k(t)$. For this reason, Thompson Sampling is called a \emph{randomized index policy}.

The TS algorithm, although being simple and easy to implement, is known to perform well for stochastic problems, for which it was proven to be asymptotically optimal \cite{AgrawalGoyal11,Kaufmann12}.
It is known to be empirically efficient, and for these reasons it has been used successfully in various applications, including on problems from Cognitive Radio \cite{Toldov,Mitton}, and also in previous work on decentralized IoT-like networks \cite{Darak16}.

\section{GNU Radio Implementation}

In this section, we present our implementation of MAB algorithms in our model of IoT networks.
We first describe the simplified physical layer of this demo,
then we present our GNU Radio implementation.

\subsection{Physical Layer and Protocol}


In this paper, we implement a PHY/MAC layers solution in order to improve the performance of IoT communications in unlicensed bands. We could have used any physical layer and any ALOHA-based protocol.
We choose to implement our own physical layer and protocol, for both clarity and conciseness.

Regarding the physical layer, we consider a QPSK constellation. Moreover, we use simplified packets composed of two parts.
The first part is the \emph{preamble} which is used for the purpose of synchronization (phase correction).
Then, we have the \emph{index} of the user, which is a sequence of QPSK symbols.
For example, this index can be a simple QPSK symbol ($\pm1\pm1j$).
Once the gateway receives an uplink packet, it detects this index and transmits an acknowledgement which has the same frame structure, but where the index is the conjugate of the index of the uplink packet (\emph{e.g.}, $1+j \mapsto 1-j$).
Thanks to this index, we can have several devices communicating with the same gateway.

In turn, the end-device that receives the acknowledgement demodulates it, and checks if the index is the conjugate of its own index.
In this case, the \emph{ACK} was for him, and it knows that its packet has been received and decoded correctly by the gateway.

\subsection{Equipment}

\begin{figure}[!t]
    \centering
    \includegraphics[width=0.95\columnwidth]{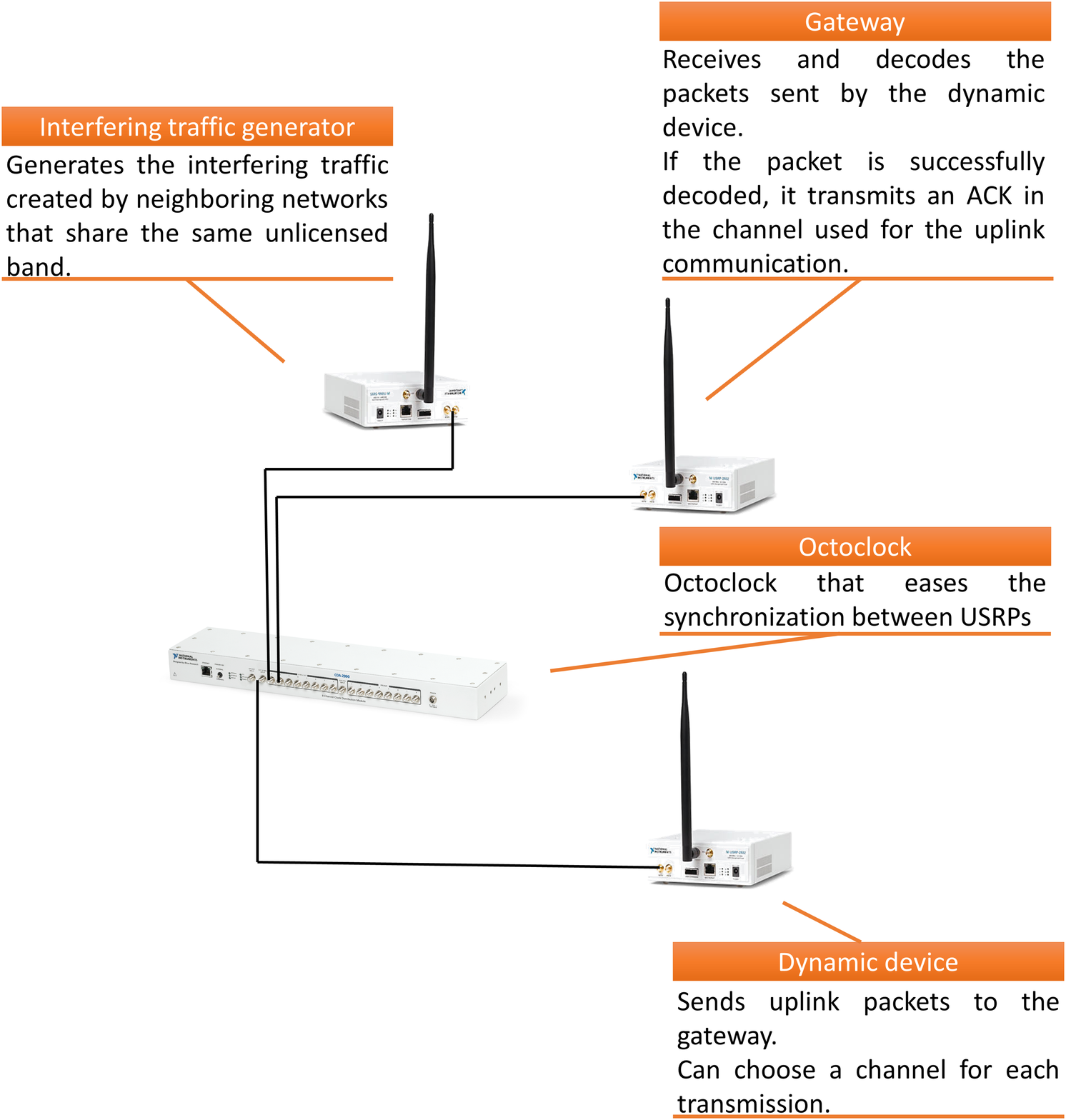}
    \caption{Schematic of our implementation that presents the role of each USRP card.}
    \label{fig:our_demo}
\end{figure}

We use USRP N210 boards \cite{USRPDocumentation}, from Ettus Research (National Instrument).
As illustrated in Figure~\ref{fig:our_demo}, our implementation is composed of at least $3$ USRP.
The gateway, a USRP which emulates the interfering traffic, and at least one dynamic device. 

The boards have their own power supply, and are all connected to a local Ethernet switch, itself connected to a single laptop, running GNU/Linux and Ubuntu.
To ease the synchronization in both time and frequency between the boards representing the dynamic objects and the gateway, we use an Octoclock \cite{OctoclockProduct}, also by Ettus Research,
and coaxial cables connecting every card to the Octoclock for time (PPS) and frequency synchronization, but this is not mandatory.

\subsection{Implementation}

We used GNU Radio Companion (GRC, version $3.7$, $2017$),
and for the demonstration the laptop runs
one GRC design to configure and control each USRP card.
As such, a single laptop can run in parallel the control program of any number of boards\footnote{~Even if in practice, maximum efficiency is kept as long as there is not more than one GRC design by CPU core.}.

GNU Radio applications are a flow-graph: a series of signal processing blocks connected together to describe a data flow.
For maximum efficiency, we wrote all of our blocks in \texttt{C++}.
GNU Radio Companion is a graphical UI used to develop GNU Radio applications:
when a flow-graph is compiled in GRC, a Python code is produced, which can be executed to connect to the USRP,
create the desired GUI windows and widgets, and create and connect the blocks in the flow-graph.

\subsection{User Interface}

We have designed a user interface in order to visualize the results obtained  with our experimental demonstration. This user interface is shown in Figure~\ref{fig:UI}.
We can see that it is made of three parts, one for each USRP, as highlighted in \textcolor{darkred}{red}:

\begin{figure*}[!t]
    \centering
    \includegraphics[width=0.68\textwidth]{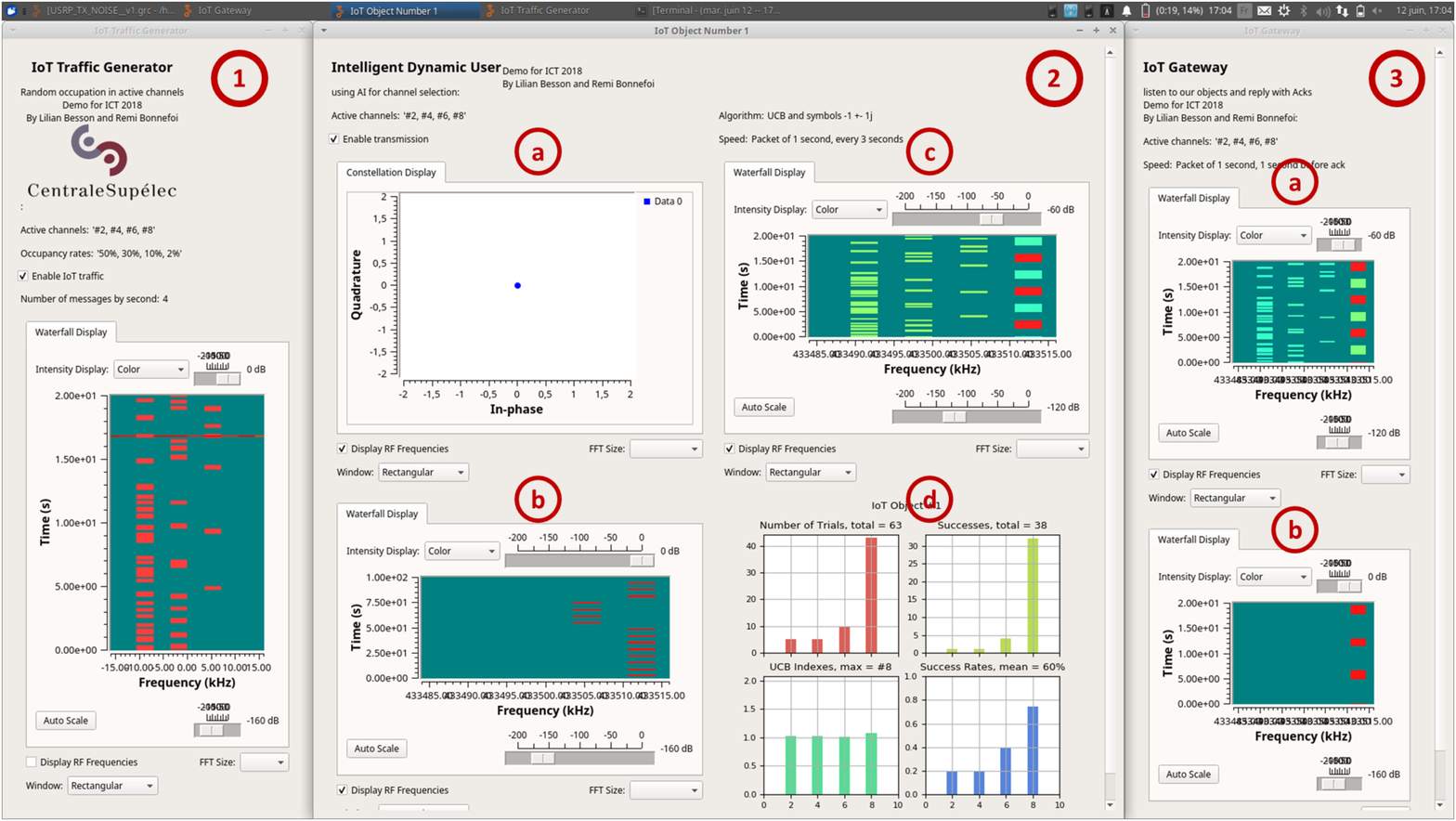}
    \caption{User interface of our demonstration.}
    \label{fig:UI}
\end{figure*}

$(1)$ The first part is the interface of the IoT traffic generator, where we see the traffic generated by this USRP, presented in a waterfall view in the time vs frequency domain.

$(2)$ The second part is the interface of the intelligent device which is made of four parts.
At the top left, we observe the constellation of the transmitted packet \emph{(a)}.
At the bottom left, we have a time/frequency view of the lasts packets transmitted by the object \emph{(b)}.
We can see, in this view that the object transmitted its last $9$ packets in the two best channels (channel $\#3$ and $\#4$).
Then, at the top right of this interface \emph{(c)}, we can see the traffic observed by this device, where we have the interfering traffic (\textcolor{darkgreen}{green}), the uplink packets transmitted by this object (\textcolor{darkred}{red}) and the acknowledgements sent by the gateway (\textcolor{darkblue}{blue}).
Finally, at the bottom right \emph{(d)}, we have four histograms showing the performance indicators of the chosen MAB algorithm (number of transmissions, number of successful transmissions, UCB indexes and success rates, in each channel).

$(3)$ The last part is the interface of the gateway, where we can see the traffic observed by the gateway \emph{(a)} and the channels in which the last acknowledgements have been sent \emph{(b)}.

\section{Results and Discussions}\label{sec:results}

\begin{figure*}[!t]
	\centering
    \includegraphics[height=8.0cm]{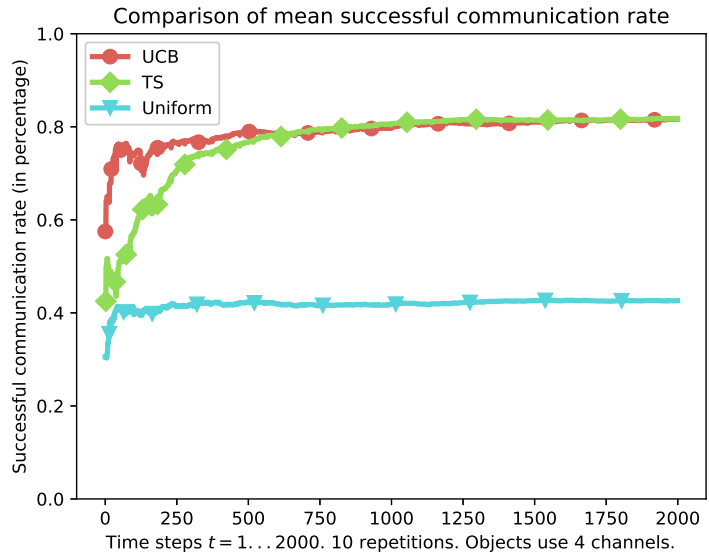}
    \caption{Less than $400$ communication slots (\emph{i.e.}, less than $100$ trials in each channel) suffice for the two learning objects to reach a successful communication rate close to $80\%$, which is \textbf{twice as much} as the non-learning (uniform) object, which stays around $40\%$ of success. Similar gains of performance were obtained in many different scenarios.}
    \label{fig:plot_datafile_append_Uniform_vs_UCB_vs_TS}
\end{figure*}

We compare the two algorithms described in Section~\ref{sub:UCB} against a uniform access algorithm, that uniformly selects its channel at random.
Three objects are compared by their mean successful communication rates, on a horizon of $2000$ communication slots, and were using three algorithms: uniform random access (in \textcolor{cyan}{cyan}), Thompson Sampling (in \textcolor{green}{green}) and \UCB{} (in \textcolor{red}{red}).
Figure~\ref{fig:plot_datafile_append_Uniform_vs_UCB_vs_TS} shows the results averaged on $10$ repetitions using the same conditions.
Each experiment takes about half a day, as we make objects generate one message every $5$ seconds, in order to artificially speed up the process and with no loss of generality.
Learning can be useful only when there is a large enough difference between ``good'' and ``bad'' channels,
Each object was learning to access $4$ different non-overlapping channels, that we chose to have occupancy rates of $[15\%, 10\%, 2\%, 1\%]$.
When facing the same stationary background traffic, we see that the learning objects are both very quickly more efficient than the naive uniform object.
We obtain an improvement in terms of successful communication rate from $40\%$ to about $60\%$ in only $100$ communications (about $16\;\mathrm{min}$), and up-to $80\%$ in only $400$ communications.
In stationary environments, both the TS and \UCB{} algorithms are very efficient and converge quickly, resulting in a very strong decrease in collisions and failed communication slots. \UCB{} is faster to learn but eventually TS gives a (slightly) better average performance.

Similar results are obtained for overlapping channels, when dynamic devices are learning in the presence of multiple devices, all using the same learning algorithm.
Empirical results confirm the simulations presented in our paper \cite[Fig.3]{Bonnefoi17}.
Such results are very encouraging, and illustrate well the various strong possibilities of MAB learning applied to IoT networks.

\section{Conclusion}


We presented in this article a demonstration,
by specifying the system model and explaining the two MAB algorithms used in practice.
We gave all the necessary details on both the PHY and the MAC layer, as well as details on the User Interface developed for the demo.
Results obtained in practice were discussed, to highlight the interest of using learning algorithms for radio online optimization problem, and especially multi-armed bandit learning algorithms.
By using such low-cost algorithms, we demonstrated empirically that a dynamically reconfigurable object can learn on its own to favor a certain channel, if the environment traffic is not uniform between the $K$ different channels.

Possible future extensions of this work include:
considering more dynamic objects (\emph{e.g.}, $100$),
implementing a real-world IoT communication protocol (like the LoRaWAN standard),
and studying the interference in case of other gateways located nearby.
%
We are also interested in studying the possible gain of using a learning step when the transmission model follows ALOHA-like retransmissions.

\subsection*{Availability of data and materials}

The source code of our demonstration is fully available online, open-sourced under GPLv3 license, at
{\small\texttt{bitbucket.org/scee\_ietr/malin-multi-arm-}}
{\small\texttt{bandit-learning-for-iot-networks-with-grc/}}.
It contains both the GNU Radio Companion flowcharts and blocks, with ready-to-use \texttt{Makefiles} to easily compile, install and launch the demonstration.

A $6$-minute \textbf{video} showing our demonstration is at \texttt{\url{youtu.be/HospLNQhcMk}}.
It shows examples of $3$ dynamic devices learning simultaneously, confirming the results of Fig.~\ref{fig:plot_datafile_append_Uniform_vs_UCB_vs_TS} for overlapping channels.

\subsection*{Acknowledgment}

The authors acknowledge the work of two CentraleSup{\'e}lec students,
Cl{\'e}ment Barras and Th{\'e}o Vanneuville, for their GNU Radio project in Spring 2017.

This work is supported by CentraleSupélec,
the French National Research Agency (ANR), under project SOGREEN (grant coded: \emph{N ANR-14-CE28-0025-02}),
R\'egion Bretagne, France,
CPER SOPHIE/STICS \& Ones,
the French Ministry of Higher Education and Research,
and ENS Paris-Saclay.


\bibliographystyle{ieeetr}
\bibliography{biblio_RIoT}


\end{document}